\documentclass[12pt]{iopart}

\newcommand{\be}{\begin{eqnarray}}
\newcommand{\ee}{\end{eqnarray}}
\newcommand{\bea}{\begin{eqnarray}}
\newcommand{\eea}{\end{eqnarray}}

\usepackage{graphicx}
\usepackage{bm}
\usepackage{color}
\usepackage{slashed}
\usepackage{cite}
\usepackage{graphicx}
\usepackage{multicol}
\usepackage{graphicx}
\usepackage{color}
\usepackage{slashed}
\usepackage{CJKutf8}
\begin{document}
\begin{CJK}{UTF8}{<font>}
\title{The shadow and photon sphere of the charged black hole in Rastall gravity}

\author{Sen Guo$^{*1}$, \ Ke-Jian He$^{2}$, \ Guan-Ru Li$^{1}$, \ Guo-Ping Li$^{*3}$}

\address{
$^1$Guangxi Key Laboratory for Relativistic Astrophysics, School of Physical Science and Technology, Guangxi University, Nanning 530004, People's Republic of China\\
$^2$College of Physics, Chongqing University, Chongqing 401331, People's Republic of China\\
$^3$Physics and Space Science College, China West Normal University, Nanchong 637000, People's Republic of China}

\ead{sguophys@126.com; gpliphys@yeah.net}
\vspace{10pt}
\begin{indented}
\item[]June 2021
\end{indented}

\begin{abstract}
Considering a charged black hole (BH) surrounded by a perfect fluid radiation field (PFRF) in Rastall gravity, we investigate this BH shadow and photon sphere on different spherical accretions backgrounds. The effect of the PFRF parameter/BH charge on the critical impact parameter is studied by investigating the light deflection near this BH. The luminosity of these BH shadows in different spherical accretions is obtained, respectively. It is found that the shadow of this BH with infalling spherical accretion is darker than static spherical accretion, and the photon sphere with infalling spherical accretion is brighter than a static one. We creatively investigate the effects of the BH charge/PFRF parameter on the luminosity of BH shadow and photon sphere. The results implying that the BH shadow is a signature of space-time geometry, and the photon sphere luminosity is affected by accretion materials and BH itself.
\end{abstract}

\noindent{\it Keywords}: Black hole shadow, Photon sphere, Spherical accretion

\section{Introduction}
\label{intro}
\par
Black hole (BH) is one of the most interesting predictions in general relativity. The gravitational wave detection result is the first strong evidence for BH existence in the universe \cite{1}. The stronger evidence is $M87^{*}$ image from the Event Horizon Telescope \cite{2}. The light ray from the universe's accretion materials is received by BH's strong gravity field, which produces a marked deficit of the observed intensity inside the apparent boundary. It is the BH shadow. Furthermore, the light ray bend causes the BH to be surrounded by shadow and make the photon sphere produces luminosity \cite{3}. Synge discovered the spherical BH shadow outline as a standard circle for the first time in $1966$ \cite{4}. Bardeen calculated the Schwarzschild BH shadow radius is $r_{s}\equiv5.2M$ \cite{5}. More researches on BH shadow appears in \cite{6,7,8,9,10,11,12,13,14,15,16,17,18,19,20}, which shows that the gravity drag of the null geodesics line cause BH shadow shape change.

\par
Apart from the study of shadow shape, the BH properties with different accretion materials have been investigated. Zeng $et.al$ found that the Gauss-Bonnet coefficient affects the Gauss-Bonnet BH shadow \cite{21}. The quintessence dark energy's contribution to BH shadow and photon ring was discussed in \cite{22}. By investigating disk accretion material, Wald $et.al$ obtained the relationship between BH shadow, photon rings, and lensing rings \cite{23}. Inspired by these researches, the luminosity of BH shadow and photon sphere on various spherical accretions backgrounds has aroused our attention. That is one of the motives in this paper.

\par
BH shadow research allows us to comprehend the BH's configuration but also help us exploring various gravity model more deeply. By investigating the charged rotating regular BHs shadow cast, Kumar $et.al$ found that BH angular momentum leads to BH shadow no longer a standard circle \cite{24}. Using the WKB approach and time-domain integration method, Konoplya $et.al$ discussed the quasinormal modes and grey-body factors with different spin and shadow cast of the quantum correction Schwarzschild solution. The results show that the radius of the shadow is decreasing when the quantum deformation is turned on \cite{25}. \"{O}vg\"{u}n $et.al$ obtained the visibility of a spherically symmetric non-commutative BH shadow depends on the non-commutative parameter in Rastall gravity \cite{26}. Then, the BH shadows characteristics have been extensively discussed on the various gravity backgrounds \cite{27,28,29,Li,He} $etc$.

\par
In another respect, Rastall proposed an extended theory of general relativity in 1972 \cite{30}. Because the current experiment cannot be sure that the derivative of energy-momentum tensor is zero in the curved space-time, Rastall thought that the equivalence principle in general relativity is stringent, who proposed the Einstein field equation is modified as $G_{\mu\nu}+\kappa \lambda g_{\mu \nu}R=\kappa T_{\mu\nu}$, the $\lambda$ is the Rastall parameter, and $\kappa\equiv{8\pi G_{N}}/{c^{4}}$ is the proportional constant in Einstein field equation that connects Einstein tensor with energy-momentum tensor \cite{30}. Interestingly, by identifying that this modified energy-momentum tensor is a physical one, Visser considers that Rastall gravity is equivalent to Einstein gravity, which is beneficial for us to have a deeper understanding of Rastall gravity \cite{Visser}. Then, Heydarzade $et.al$ investigated solutions of BH surrounded by perfect fluid in Rastall theory \cite{31}, and Pourhassan $et.al$ obtained the thermodynamics of these BH solutions \cite{32}.

\par
Nevertheless, the shadow and photon sphere of the charged BH surrounded by PFRF in Rastall theory contexts research is still of opening question. This paper focuses on this issue. We focus primarily on the simple case of emission from an optically and geometrically thin spherical accretion near this BH. Considering the static and infalling spherical accretions, we analyze this BH shadow and photon sphere with spherical accretions and further investigate the influence of BH parameters on their luminosity.

\par
The organization of this work is as follows. In Section \ref{sec:2}, a charged BH surrounded by PFRF in Rastall theory is considered, and we investigate the light deflection near this BH. In Section \ref{sec:3}, we study the shadow and photon sphere of this BH in various spherical accretions. We draw the conclusions and discussions in Section \ref{sec:4}. For simplicity, we adopt the units $G_{N}=\hbar=\kappa_{B}=c=1$.

\section{The charged BH surrounded by PFRF in Rastall gravity and light deflection}
\label{sec:2}
\par
Considering a charged BH surrounded by PFRF in Rastall theory, the BH metric is given by \cite{31}
\begin{equation}
\label{2-1}
ds^{2}=-f(r)dt^{2}+\frac{dr^{2}}{f(r)}+r^{2}d\Omega^{2},
\end{equation}
where BH metric potential can be written as
\begin{equation}
\label{2-2}
f(r)=1-\frac{2M}{r}+\frac{4\pi G_{N} (Q^{2}-N_{r})}{c^{4}r^{2}},
\end{equation}
in which $M$ is BH mass, $Q$ is BH charge, and $N_{r}$ is the radiation field (RF) parameter. We found that this metric potential is similar to the Reissner-Nordstr\"{o}m (RN) BH with an effective charge $Q_{eff}=\sqrt{Q^{2}-N_{r}}$. Compared to the RN BH, the appearance of effective charge in the BH solution non-change the causal structure and Penrose diagrams of this BH solution. Taking the natural unit, the charged BH surrounded by PFRF in Rastall gravity horizon radius is obtained, i.e.
\begin{equation}
\label{2-3}
r^{P}_{\pm}=M\pm \sqrt{M^{2}-Q^{2}+N_{r}},
\end{equation}
where $r^{P}_{\pm}$ represents event horizon radius ($r^{P}_{+}$) and inner horizon radius ($r^{P}_{-}$), respectively. Compared with the horizon radius of RN BH, our BH contain RF parameter term, this result is interpreted as the positive contribution of the RF characteristic to BH effective charge.

\par
Next, we reviewed the deflection of photons. Due to the interaction between light ray and BH gravity field, the light rays deflect when it passes from the BH vicinity. The motion of light rays satisfies the Euler-Lagrangian equation in space-time. We have
\begin{equation}
\label{2-4}
\frac{d}{d\zeta}\Big(\frac{\partial \mathcal{L}}{\partial \dot{x}^{\alpha}}\Big)=\frac{\partial \mathcal{L}}{\partial x^{\alpha}},
\end{equation}
where $\zeta$ is an affine parameter, $\dot{x}^{\alpha}$ is the four-velocity of photon. The Lagrangian $\mathcal{L}$ including the zero geodesics as following
\begin{equation}
\label{2-5}
\mathcal{L}=-\frac{1}{2}g_{\alpha \beta}\dot{x}^{\alpha}\dot{x}^{\beta}=\frac{1}{2}\Big(f(r)\dot{t}^{2}-\frac{\dot{r}^{2}}{f(r)}-r^{2}\dot{\theta}^{2}-r^{2}\sin^{2}\theta \dot{\varphi}^{2}\Big).
\end{equation}
Because of the symmetry, it suffices to consider geodesics in the equatorial plane, i.e. $\theta_{0}=\pi/2$, $\dot{\theta_{0}}=0$, and $\ddot{\theta_{0}}=0$. According to equations (\ref{2-2}) and (\ref{2-5}), BH metric is static spherically symmetric, which implies that it is a non-function of time $t$ and angle $\varphi$, satisfying $({\partial \mathcal{L}}/{\partial t})=0$ and $({\partial \mathcal{L}}/{\partial \varphi})=0$. Therefore, there is a pair of conserved quantities of energy and angular momentum
\begin{eqnarray}
\label{2-6}
E=\Big(\frac{\partial \mathcal{L}}{\partial \dot{t}}\Big)=f(r)\frac{dt}{d\zeta},~~~~~~~~~L=-\Big(\frac{\partial \mathcal{L}}{\partial \dot{\varphi}}\Big)=r^{2} \frac{d\varphi}{d\zeta}.
\end{eqnarray}

\par
The four-velocity of time, azimuthal, and radial components satisfy the motion equation,
\begin{eqnarray}
\label{2-7}
\Big(\frac{dt}{d\zeta}\Big)=\frac{L}{bf(r)},~~~~~~\Big(\frac{d\varphi}{d\zeta}\Big)=\pm\frac{L}{r^{2}},~~~~~~\Big(\frac{dr}{d\zeta}\Big)^{2}+\frac{f(r)}{r^{2}}=\frac{1}{b^{2}},
\end{eqnarray}
where $b$ is defined as impact parameter ($b\equiv|L|/E$) and $\pm$ corresponding to the light rays motion counterclockwise and clockwise direction, respectively. Considering the null geodesic $\mathcal{L}=0$, the orbit equation $(\dot{r}^{2}/\dot{\varphi}^{2})$ for lightlike geodesics is obtained. Hence, the effective potential of the charged BH surrounded by PFRF in Rastall gravity can be written as
\begin{equation}
\label{2-8}
V_{eff}=\frac{f(r)}{r^{2}}=\frac{1}{r^{2}}\Big(1-\frac{2M}{r}+\frac{Q^{2}-N_{r}}{r^{2}}\Big).
\end{equation}
The photon sphere orbit conditions are $\dot{r}=0$ and $\ddot{r}=0$, the effective potential should satisfy the conditions
\begin{equation}
\label{2-9}
V_{eff}=\frac{1}{b^{2}},~~~~~~~~~V'_{eff}=0.
\end{equation}

\par
Based on equation (\ref{2-9}), the numerical results of the event horizon radius $r^{P}_{+}$ of this BH, the shadow radius $r^{P}_{sh}$, and impact parameter $b_{ph}$ of photon sphere for different values of BH charges/RF parameters have been given in Table 1 and Table 2. We can see that $r^{P}_{+}$, $r^{P}_{sh}$ and $b_{ph}$ are all smaller and smaller with the increase of the BH charge when $N_{r}$ is given. Moreover, the parameter $N_{r}$ has the opposite effect. It is worth mentioning that we adopt the natural unit system for simplicity, thus giving the pure number solution in tables.
\begin{center}
\label{table:1}
{\footnotesize{\bf Table 1.} The event horizon, shadow radius and impact parameter for different $Q$ with $M=1$ and $N_{r}=1$.\\
\vspace{2mm}
\begin{tabular}{ccccccc}
\hline
{Q} & {0.2} & {0.4} & {0.55} & {0.7} & {0.85} & {0.9}\\\hline
$r^{P}_{+}$   &  $2.4$      & $2.35647$  &  $2.30288$  &  $2.22882$  &  $2.13027$  &  $2.09087$ \\
$r^{P}_{sh}$  &  $3.54206$  & $3.48242$  &  $3.40919$  &  $3.30831$  &  $3.17481$  &  $3.12173$ \\
$b_{ph}$      &  $5.91297$  & $5.8331$   &  $5.7353$   &  $5.60116$  &  $5.42477$  &  $5.35504$ \\
\hline
\end{tabular}}
\end{center}
\begin{center}
\label{table:2}
{\footnotesize{\bf Table 2.} The event horizon, shadow radius and impact parameter for different $N_{r}$ with $M=1$ and $Q=0.6$.\\
\vspace{2mm}
\begin{tabular}{ccccccc}
\hline
{RF} & {1} & {2} & {3} & {4} & {5.5} & {7}\\\hline
$r^{P}_{+}$   &  $2.28062$   & $2.62481$  & $2.90788$  &  $3.37487$  &  $3.4779$   &  $3.76405$ \\
$r^{P}_{sh}$  &  $3.378832$  & $3.8516$   & $4.24408$  &  $4.89559$  &  $5.03977$  &  $5.44081$ \\
$b_{ph}$      &  $5.69486$   & $6.33041$  & $6.86507$  &  $7.76132$  &  $7.96073$  &  $8.51686$ \\
\hline
\end{tabular}}
\end{center}

\par
We plot the effective potential as a function of $r$ for different BH charges/RF parameters in figure 1. It is evident that the radius and critical impact parameters decrease with an increase in BH charge for a given RF parameter. In contrast, the RF parameter has the opposite effect when the BH charge is fixed.
\begin{center}
\includegraphics[width=6cm,height=5cm]{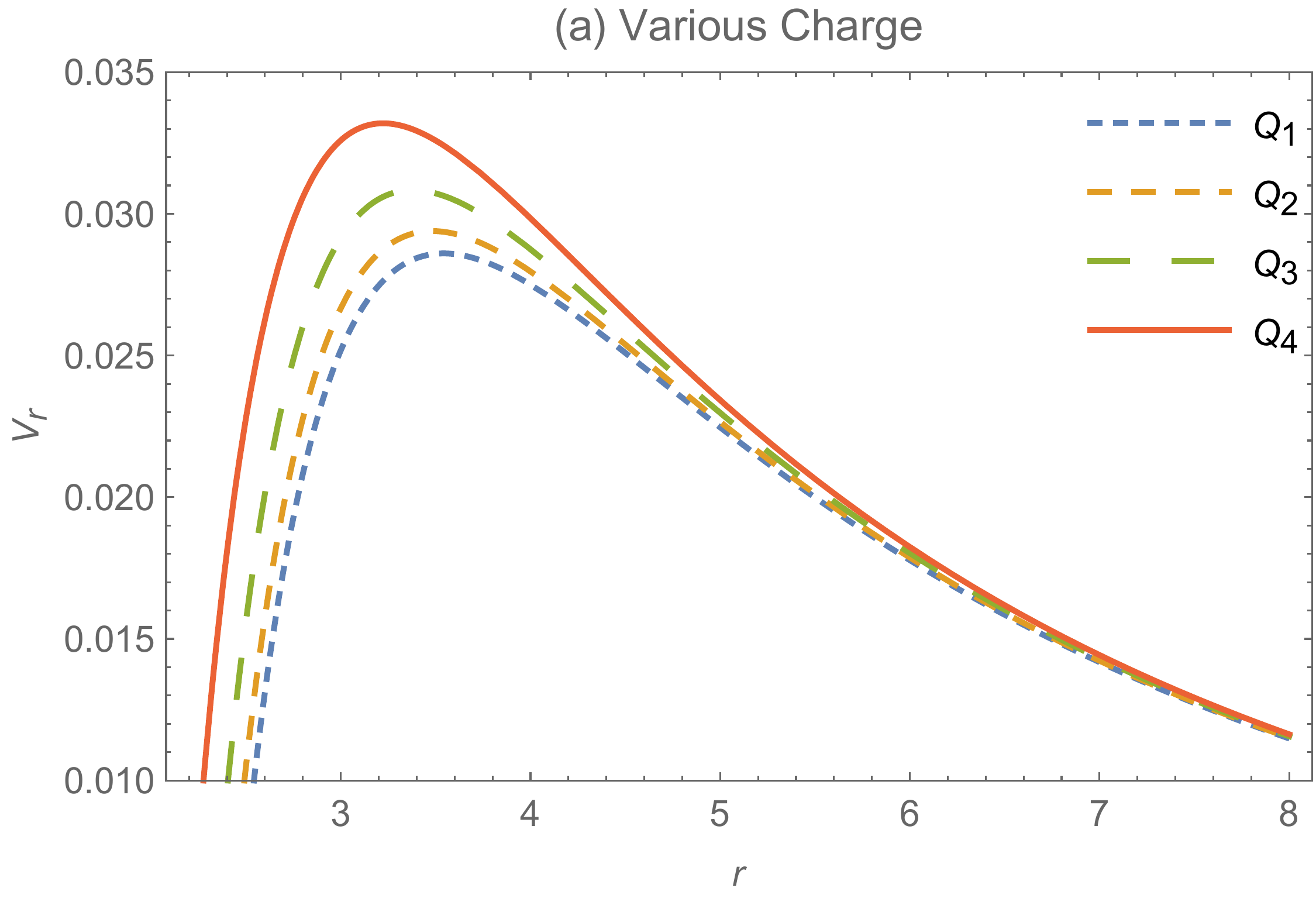}
\includegraphics[width=6cm,height=5cm]{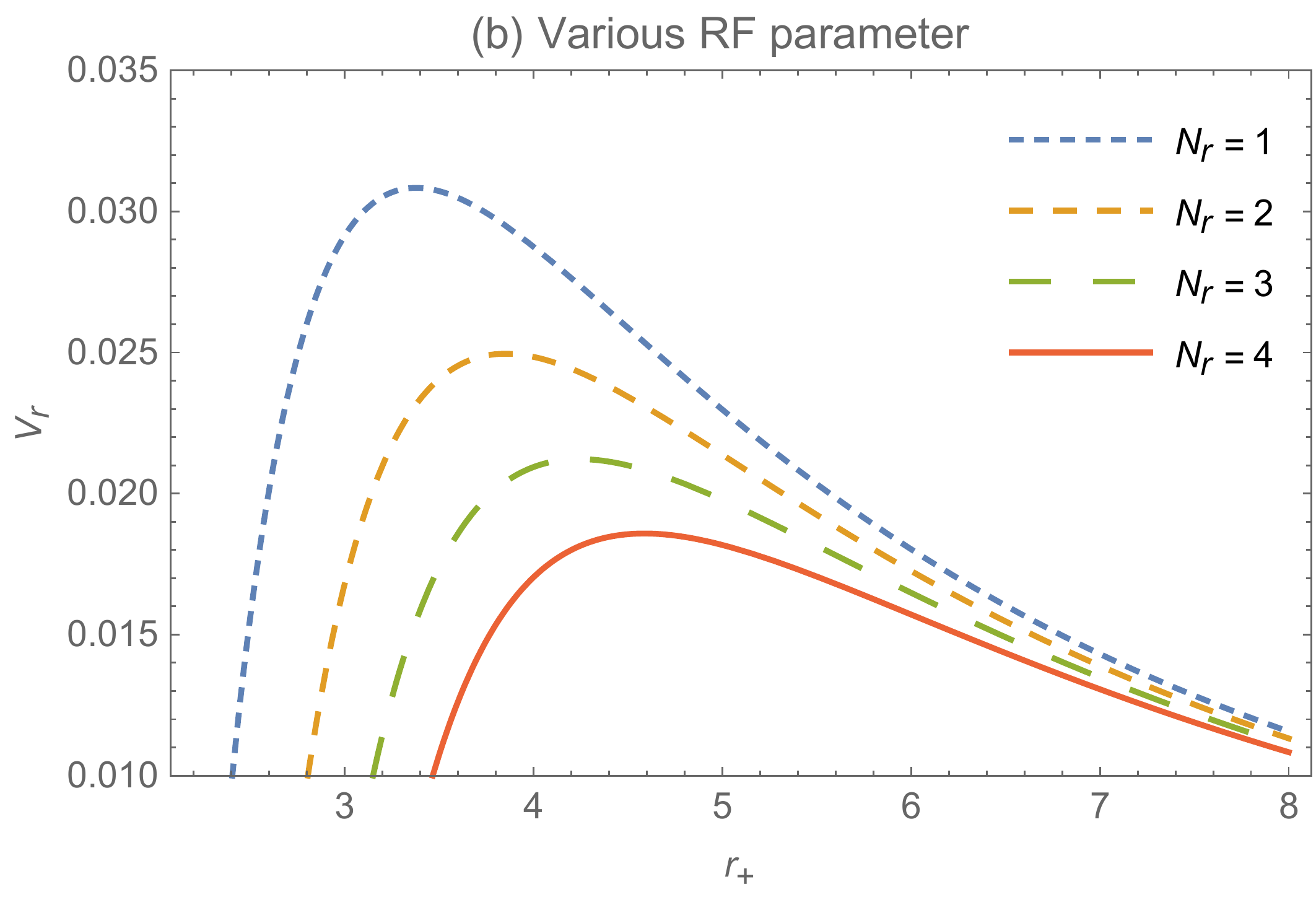}
\parbox[c]{15.0cm}{\footnotesize{\bf Fig~1.}  
Effective potential $V(r)_{eff}$ as a function of radius $r$. {\em Panel (a)} -- Various charges $Q=0.2,0.4,0.6,0.8$ with $N_{r}=1, M=1$ and {\em Panel (b)} -- Various RF parameters $N_{r}=1,2,3,4$ with $Q=0.6, M=1$.}
\label{fig1}
\end{center}

\par
By introducing a new parameter $u\equiv1/r$ and according to equation (\ref{2-7}), we have
\begin{equation}
\label{2-10}
\frac{du}{d\varphi}=\sqrt{\frac{1}{b^{2}}-u^{2}(1-2M u- N_{r}u^{2}+u^{2}Q^{2})}\equiv \Omega(u).
\end{equation}
Assuming that all the light rays from accretions approach the BH from the right side, the BH produces shadow as the light is deflected. By using the ray-tracing code \cite{33} and equation (\ref{2-10}), the light ray trajectory is shown in figure 2.
\begin{center}
\includegraphics[width=5.5cm,height=5.5cm]{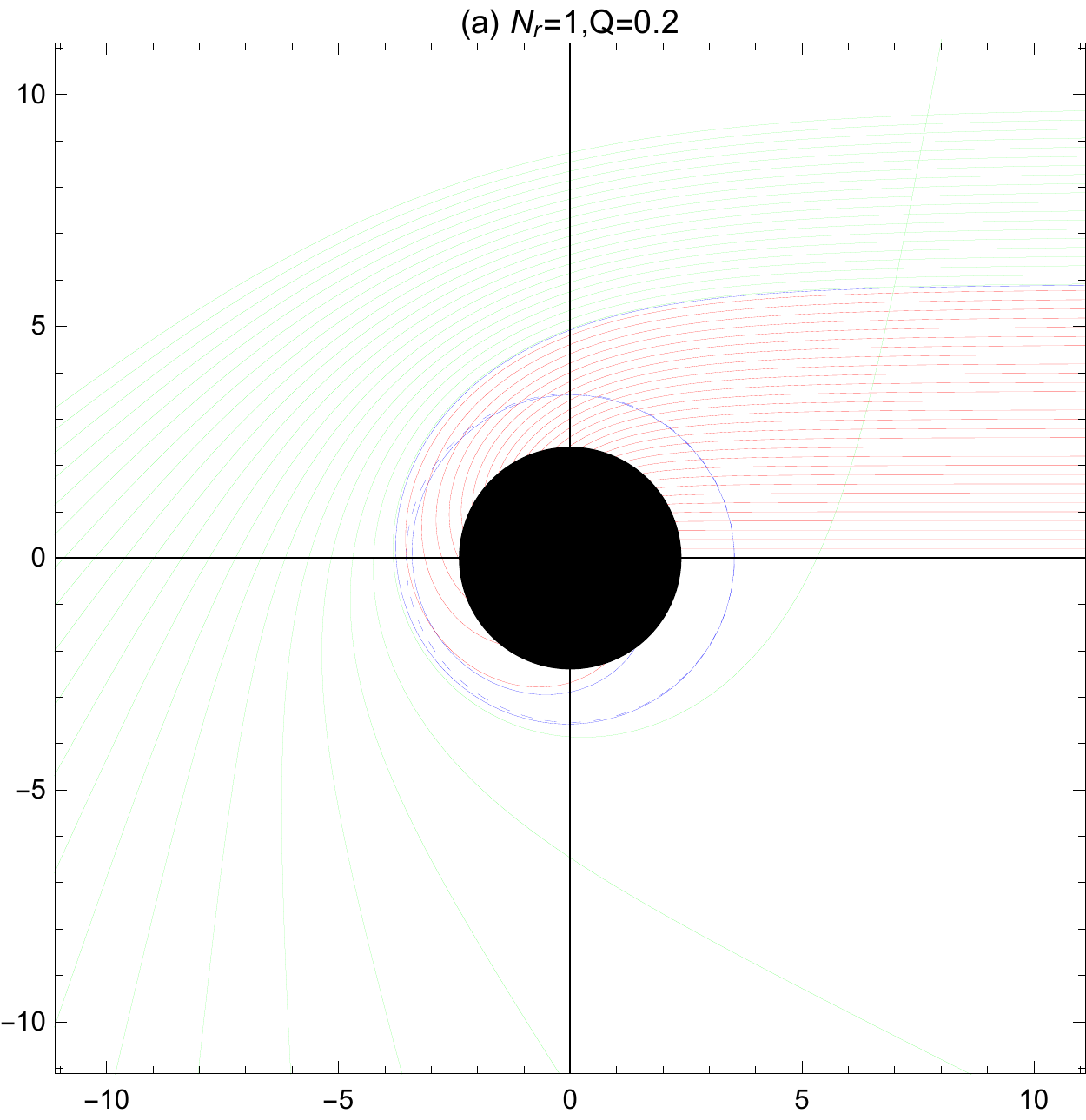}
\includegraphics[width=5.5cm,height=5.5cm]{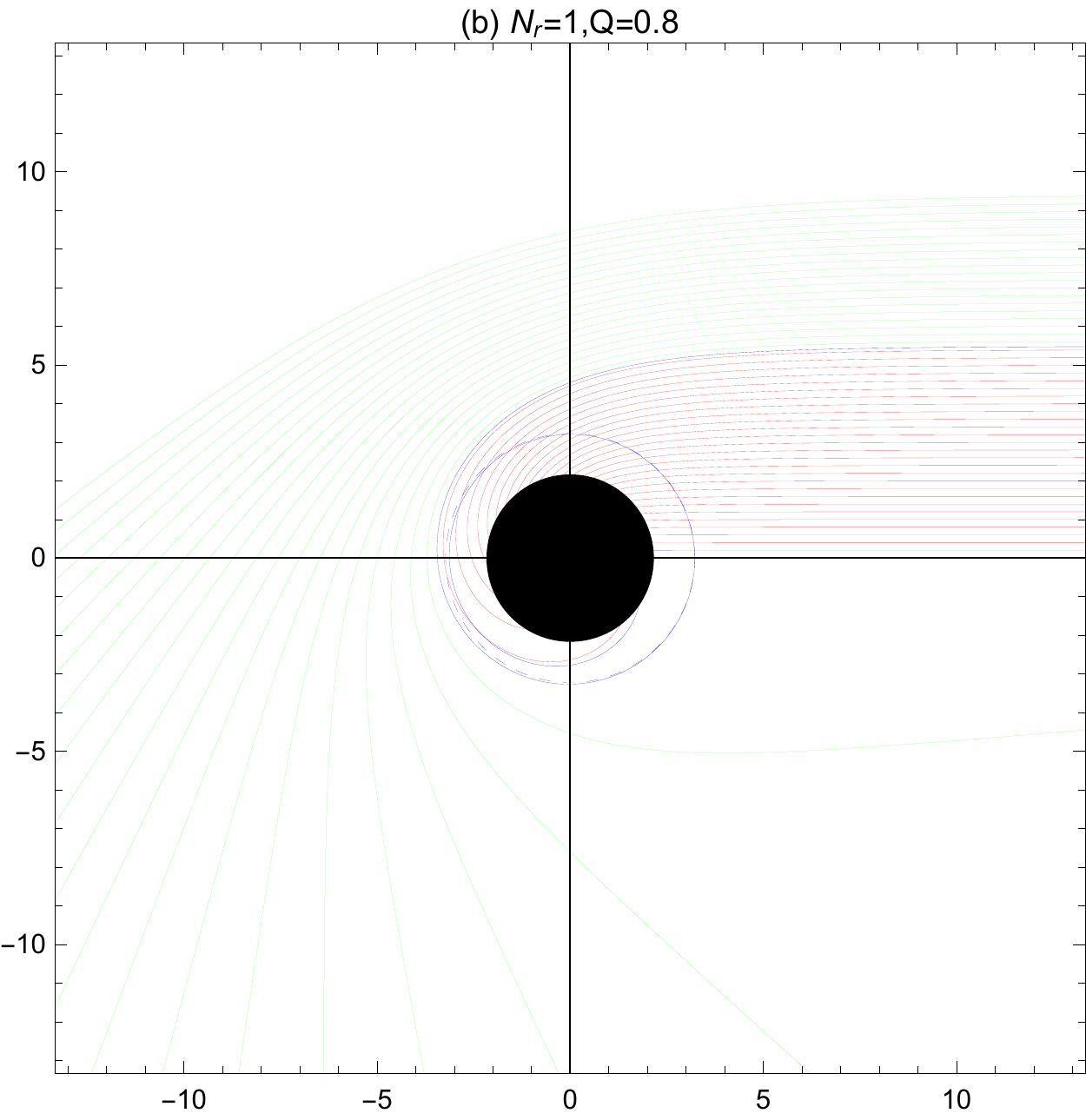}
\includegraphics[width=5.5cm,height=5.5cm]{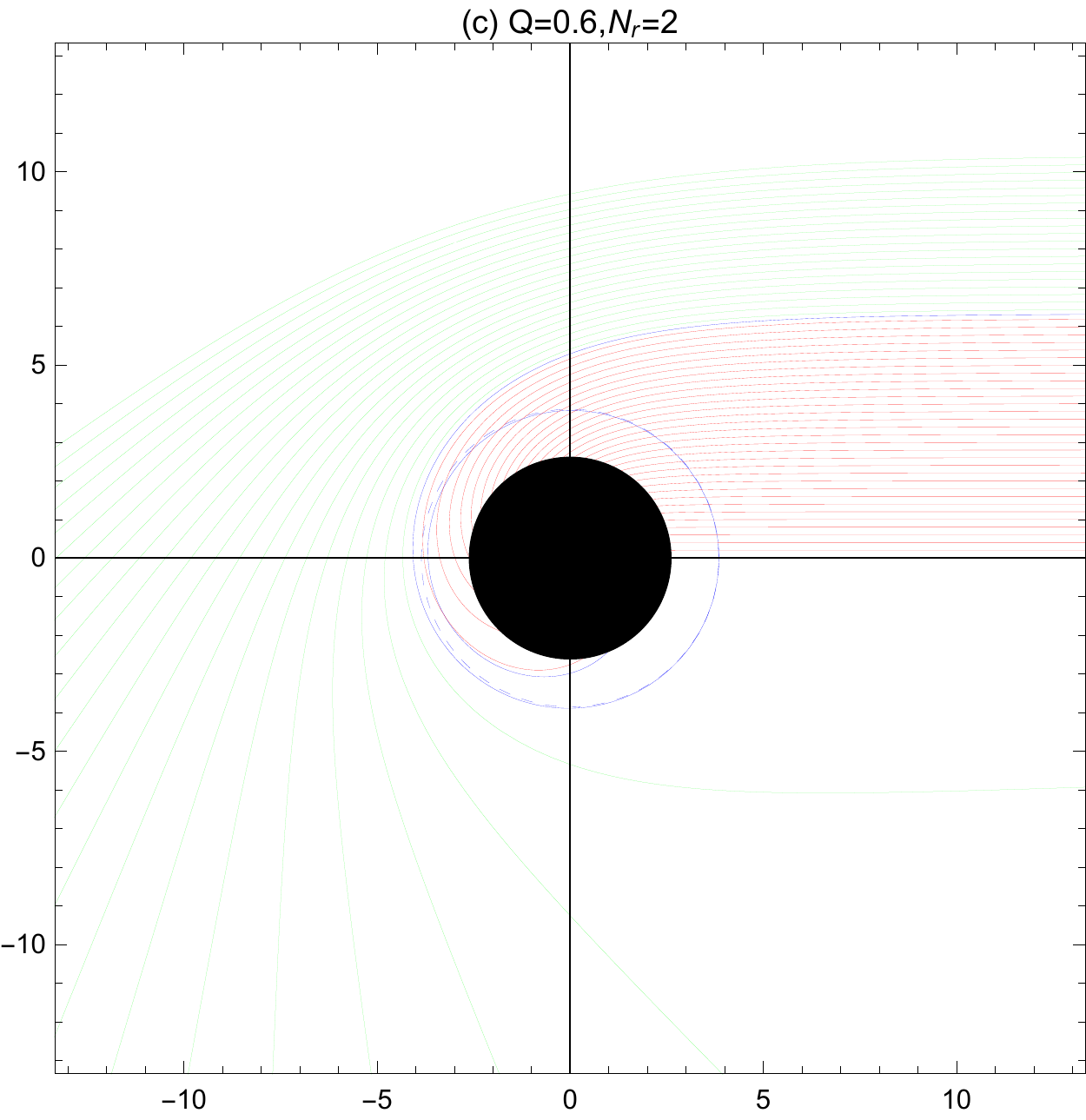}
\includegraphics[width=5.5cm,height=5.5cm]{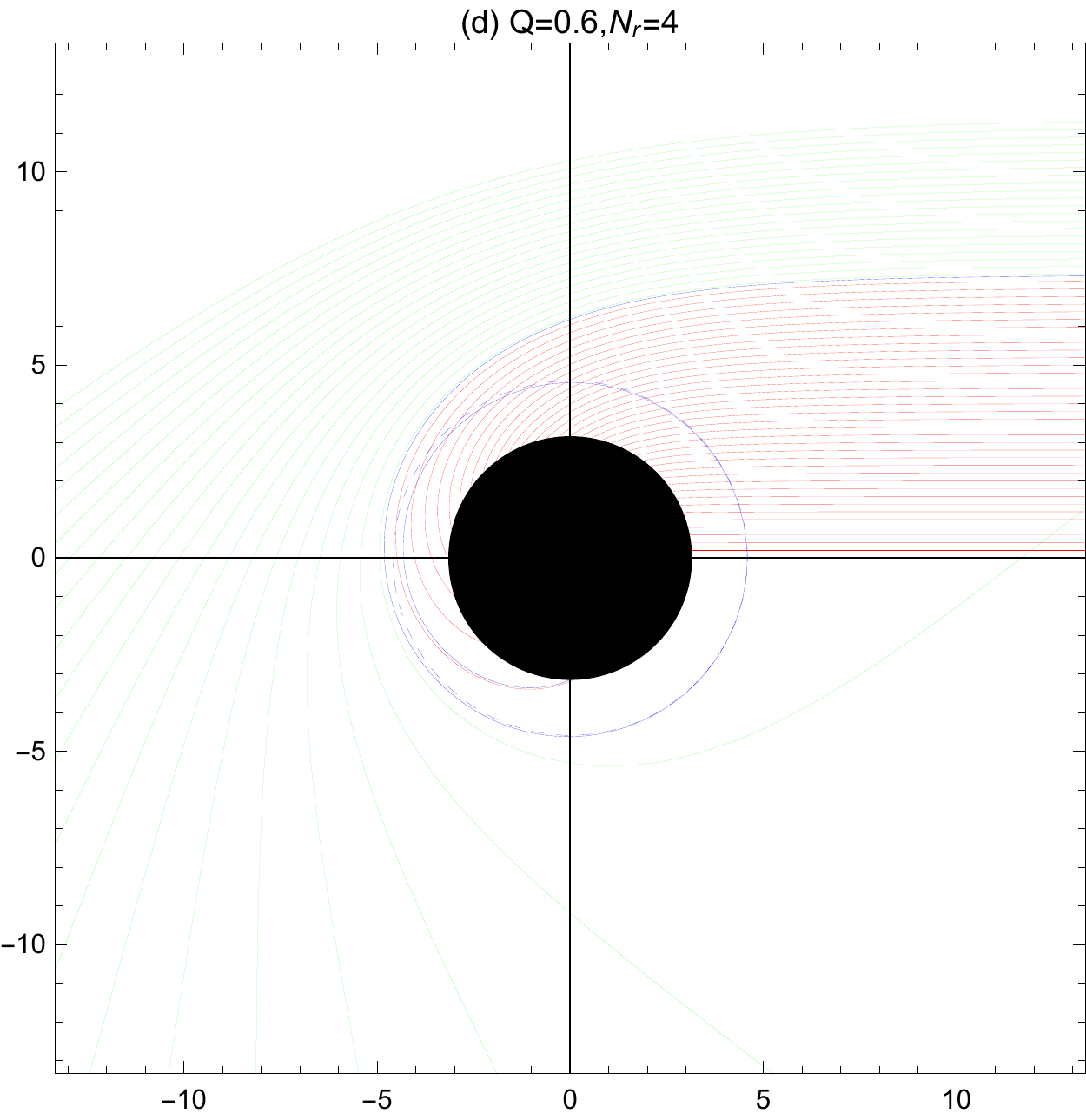}
\parbox[c]{15.0cm}{\footnotesize{\bf Fig~2.}  
{\em Panel (a, b)} -- Various charges $Q=0.2,0.8$ with $N_{r}=1, M=1$. and {\em Panel (c, d)} -- Various RF parameters $N_{r}=2,4$ with $Q=0.6, M=1$. The green lines, blue lines and red lines correspond to $b>b_{ph}$, $b=b_{ph}$ and $b<b_{ph}$, respectively. The BH is shown as a solid disk and photon orbit as a dashed blue line.}
\label{fig2}
\end{center}

\par
In figure 2, it is evident that this BH shadow radius $r^{P}_{sh}$ decreases when the BH charge increases. In comparison, the BH shadow radius increases with an increase in the RF parameter for given BH charge, which is consistent with the cases in table 1 and table 2. Compare the results of figure 2--(a) and (b), we found that the deflection of a light ray at the shadow is smaller, and the light ray density is lower when the value of $q$ is larger. By comparing figure 2--(c) and (d), the deflect light ray at the shadow is larger, and the light ray density is higher when the value of $N_{r}$ is larger.

\section{BH shadow and photon sphere on different spherical accretions}
\label{sec:3}
\subsection{The static spherical accretion}
\label{sec:3-1}
\par
Considering static spherical accretion for the thin of optically and geometrically. The intensity of radiation $I(\upsilon_{obs})$ coming to a far observer at the frequency $\upsilon_{obs}$ along any ray is given by \cite{34,35}
\begin{equation}
\label{3-1-1}
I(\upsilon_{obs})=\int_{ray}g^{3}j(\upsilon_{em})dl_{prop},
\end{equation}
where $g\equiv \upsilon_{obs}/\upsilon_{em}$ is gravity redshift factor, $\upsilon_{em}$ is emitter photon frequency, $j(\upsilon_{em})$ is emissivity per unit volume, and $dl_{prop}$ is proper length differential as measured in the frame comoving with the matter. By assuming monochromatic with rest-frame frequency $\upsilon_{t}$, the proper length measured in the rest frame of emitter for BH is obtained, i.e.
\begin{eqnarray}
\label{3-1-2}
g=f(r)^{1/2},~~j(\upsilon_{em})\propto \frac{\delta(\upsilon_{em}-\upsilon_{t})}{r^2},~~dl_{prop}=\sqrt{f(r)^{-1}+r^{2}\Big(\frac{d\varphi}{dr}\Big)^{2}}dr.
\end{eqnarray}
According to equations (\ref{3-1-1}) and (\ref{3-1-2}), the specific intensity observed $I(\upsilon_{obs})$ by the distant observer can be re-written as
\begin{equation}
\label{3-1-3}
I(\upsilon_{obs})=\int_{ray}\frac{f(r)^{3/2}}{r^{2}}\sqrt{f(r)^{-1}+r^{2}\Big(\frac{d\varphi}{dr}\Big)^{2}}dr.
\end{equation}

\par
Based on equations (\ref{2-2}) and (\ref{3-1-3}), the luminosity of this BH shadow and photon sphere on static spherical accretion background for a distant observer is obtained. We plot specific intensity $I(\upsilon_{obs})$ as a function of impact parameter $b$ for different RF parameters and BH charges in figure 3, respectively. We can see that $I(\upsilon_{obs})$ reaches a peak rapidly with the increased impact parameter and decreases gradually to a minimum value. Because the light rotates around BH many times in photon sphere orbit, the optical path is infinite, and the resulting $I(\upsilon_{obs})$ is the strongest at $b_{ph}$. In addition, the peak value of intensity increase with an increase of BH charge when the RF parameter is a constant, but $b$ corresponding to the peak value decreases with an increase of BH charge. The result is consistent with the results in table 1. The RF parameter has the opposite effect when the BH charge is fixed.
\begin{center}
\includegraphics[width=6cm,height=5cm]{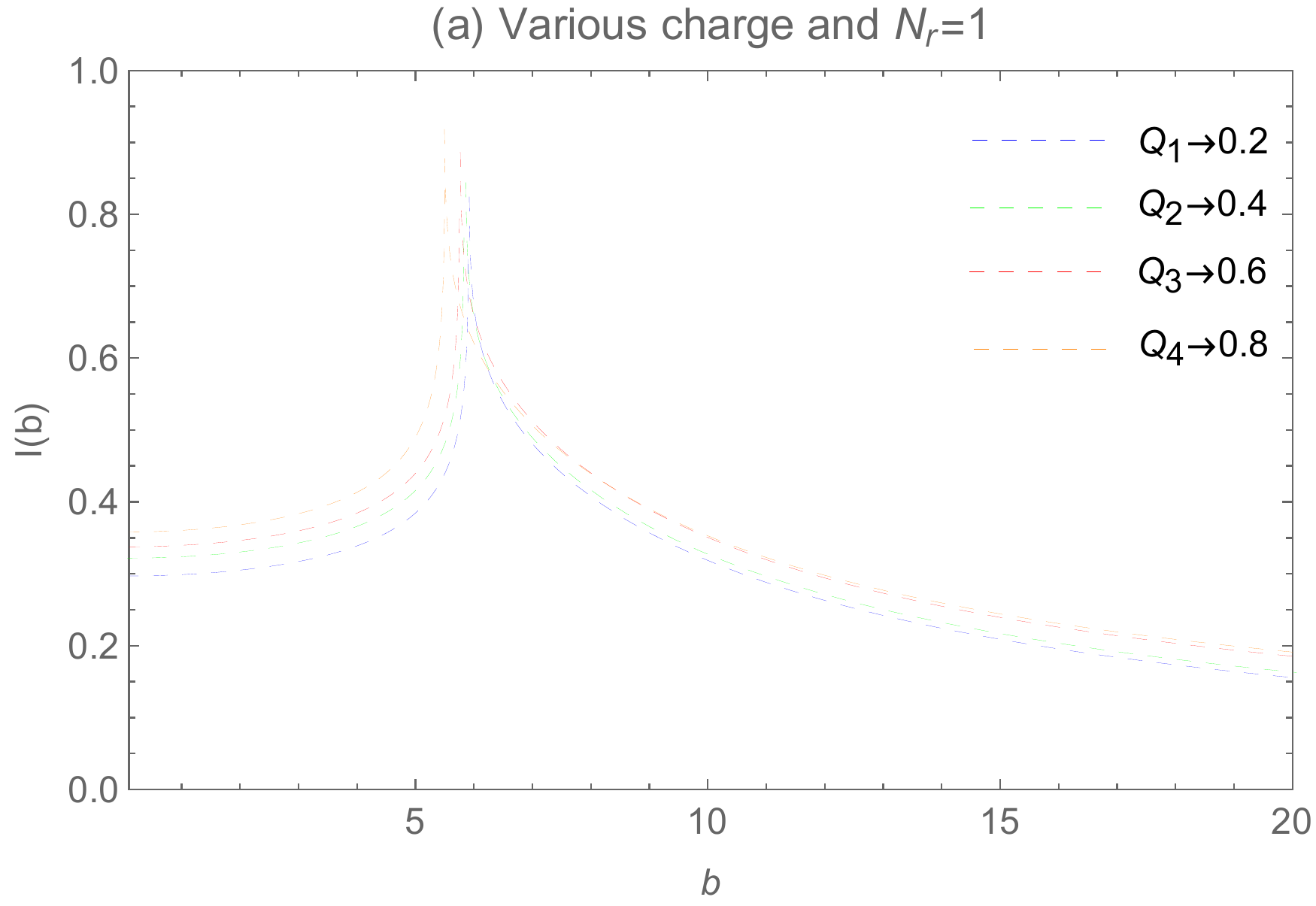}
\includegraphics[width=6cm,height=5cm]{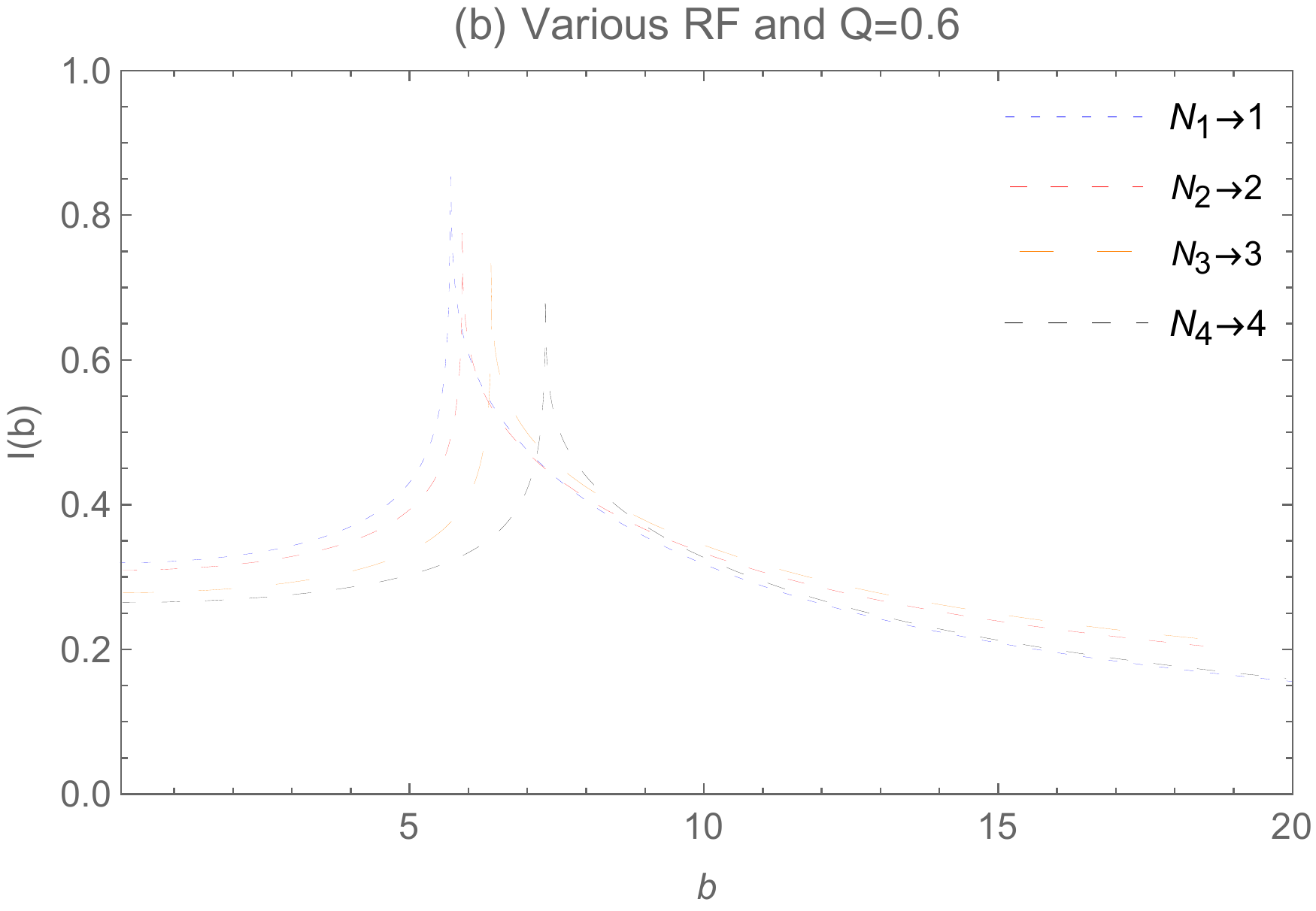}
\parbox[c]{15.0cm}{\footnotesize{\bf Fig~3.}  
Profiles of intensity $I(b)$ with static spherical accretion. {\em Panel (a)} -- Various charges $Q=0.2,0.4,0.6,0.8$ with $N_{r}=1, M=1$ and {\em Panel (b)} -- Various RF parameters $N_{r}=1,2,3,4$ with $Q=0.6, M=1$.}
\label{fig3}
\end{center}

\par
Furthermore, the shadow cast by this BH in the $(x,y)$ plane is shown in figure 4. We can see that the photon sphere with the strongest luminosity is outside of the BH shadow, and the inner region of the shadow is not entirely black. Because of the radiation field existence, the tiny fraction of photons escape from BH, and the photon sphere luminosity becomes more prominent with the enhancement of the RF parameter. Moreover, the maximum luminosity of the shadow image is attenuated with the increase of the BH charge when the RF parameter is fixed.
\begin{center}
\includegraphics[width=6cm,height=6cm]{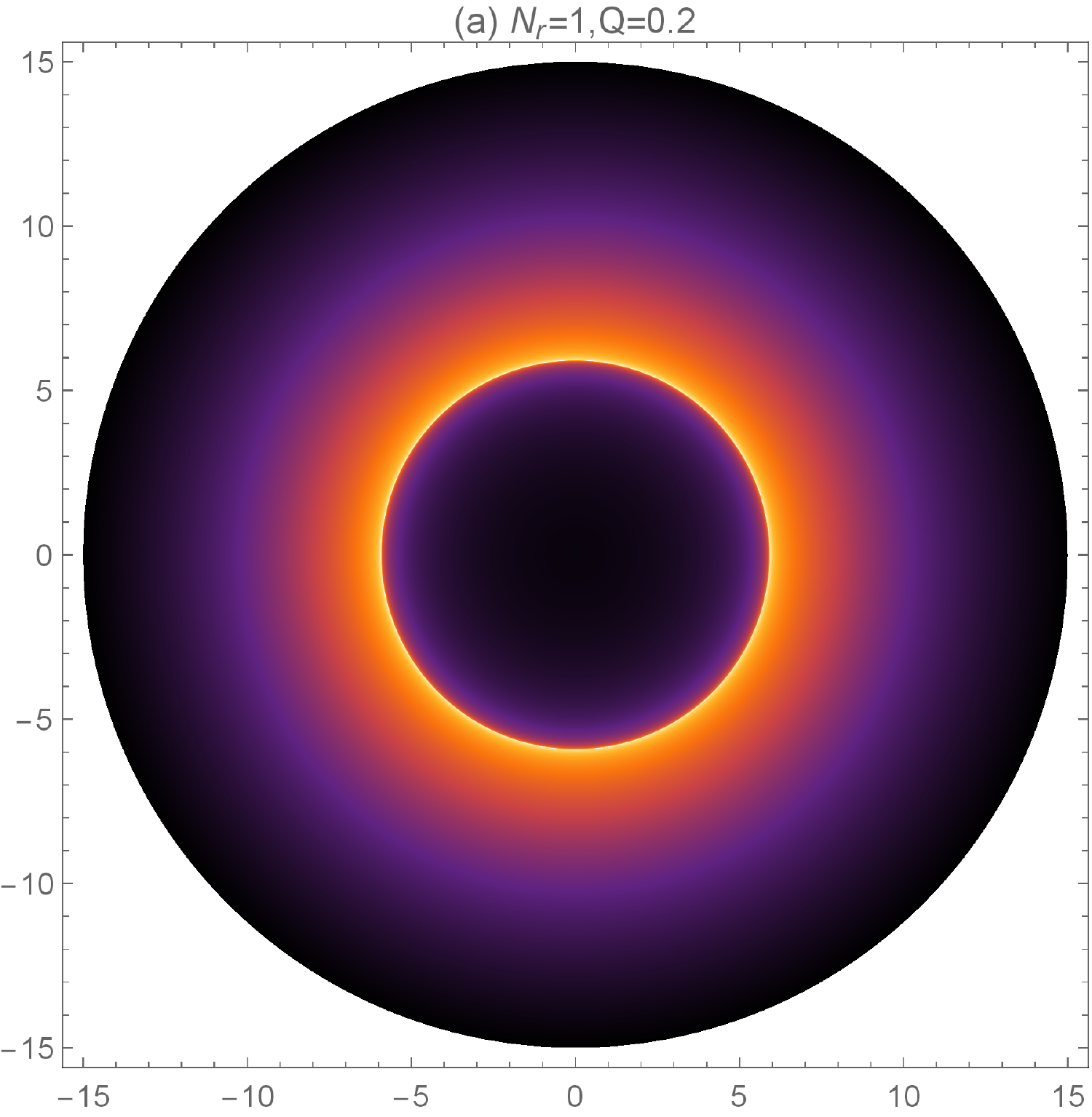}
\includegraphics[width=0.4cm,height=6cm]{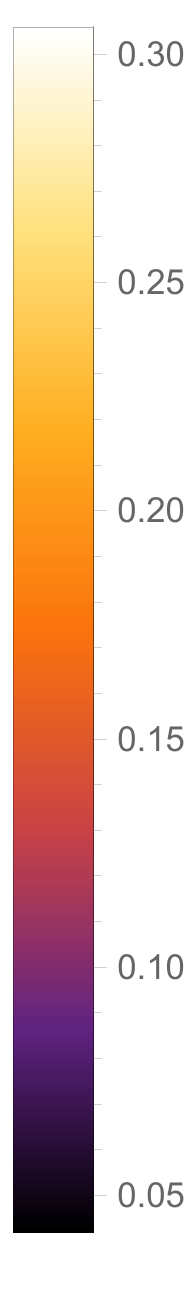}
\includegraphics[width=6cm,height=6cm]{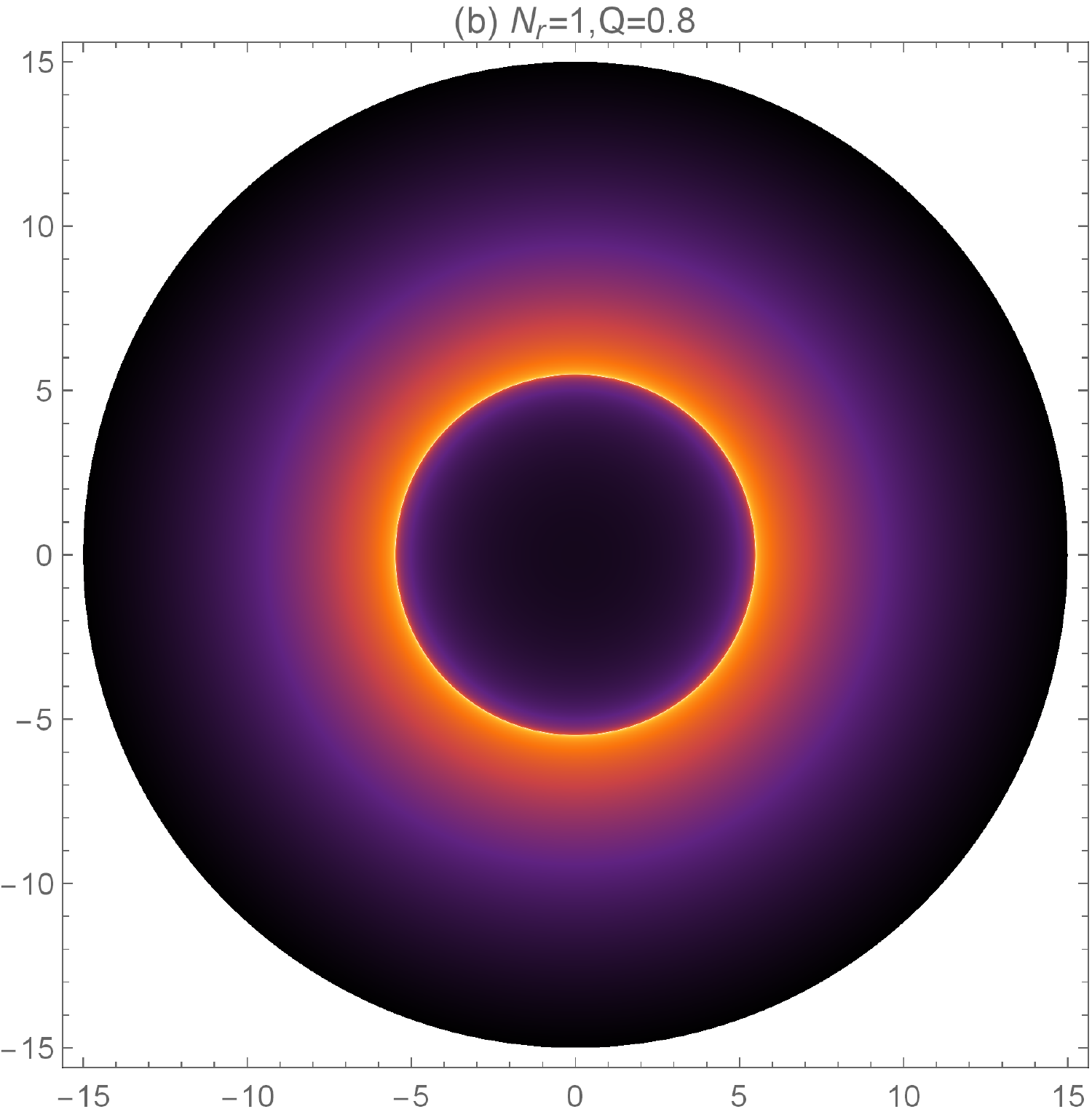}
\includegraphics[width=0.4cm,height=6cm]{fig0.pdf}
\includegraphics[width=6cm,height=6cm]{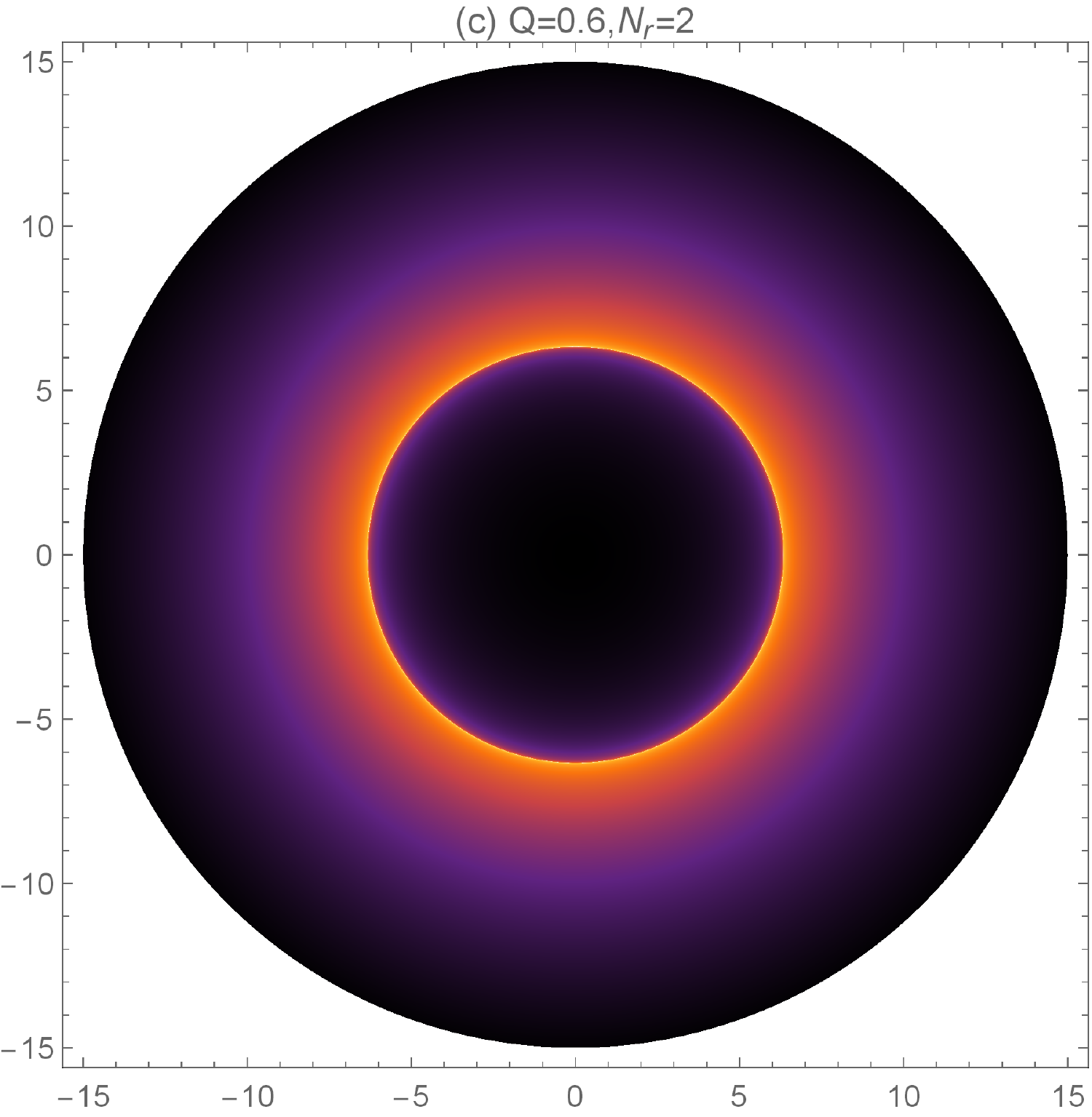}
\includegraphics[width=0.4cm,height=6cm]{fig0.pdf}
\includegraphics[width=6cm,height=6cm]{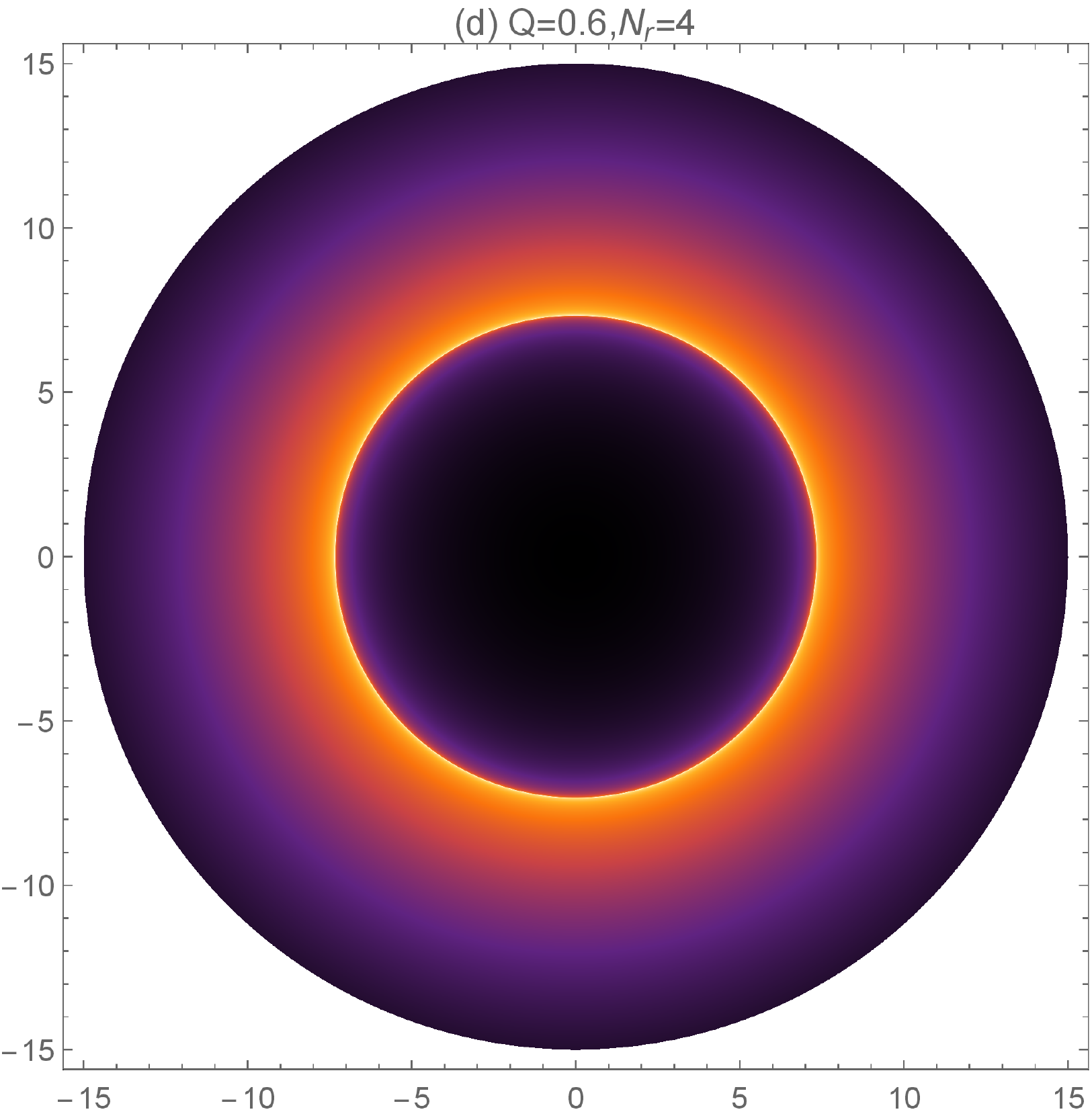}
\includegraphics[width=0.4cm,height=6cm]{fig0.pdf}
\parbox[c]{15.0cm}{\footnotesize{\bf Fig~4.}  
BH shadows and photon spheres cast with static spherical accretion in the (x,y) plane. {\em Panel (a, b)} -- Various charges $Q=0.2,0.8$ with $N_{r}=1, M=1$ and {\em Panel (c, d)} -- Various RF parameters $N_{r}=2,4$ with $Q=0.6, M=1$.}
\label{fig4}
\end{center}

\subsection{The infalling spherical accretion}
\label{sec:3-2}
\par
As we all know, most accretions are moving in the universe. This section considers BH shadow and photon sphere on infalling spherical accretion background, which is more accurate than static spherical accretion. Assuming that infalling spherical accretion falling onto BH from infinity, equation (\ref{3-1-3}) is still determined to be effective. The redshift factor, in this case, is different from static spherical accretion. It is
\begin{equation}
\label{3-2-1}
g=\frac{k_{\alpha}u_{obs}^{\alpha}}{k_{\vartheta}u_{em}^{\vartheta}},
\end{equation}
where $k^{\mu}\equiv \dot{x}_{\mu}$, $u_{obs}^{\mu}\equiv(1,0,0,0)$, $u_{em}^{\mu}$ corresponding to photon four-velocity, distant observer four-velocity and accretion four-velocity, respectively. According to equation (\ref{2-7}), one can get $k_{t}$ is a constant $(1/b)$, and $k_{r}$ comes from $k_{\vartheta}k^{\vartheta}=0$, hence, we have
\begin{equation}
\label{3-2-2}
\frac{k_{r}}{k_{t}}=\pm\frac{1}{f(r)}\sqrt{1-\frac{b^{2}f(r)}{r^{2}}},
\end{equation}
where the sign here indicates that photons are approaching or away from BH. The accretion under consideration four-velocity as
\begin{eqnarray}
\label{3-2-3}
u_{em}^{t}=\frac{1}{f(r)},~~~~~u_{em}^{\theta}=u_{em}^{\varphi}=0,~~~~~u_{em}^{r}=-\sqrt{1-f(r)}.
\end{eqnarray}
According to above equations, the redshift factor of infalling spherical accretion can be written as
\begin{equation}
\label{3-2-4}
g_{f}=\frac{1}{u_{em}^{t}+k_{r}/k_{em}u_{em}^{r}},
\end{equation}
and the proper distance as
\begin{equation}
\label{3-2-5}
dl_{prop}=k_{\vartheta}u_{em}^{\vartheta}d\zeta=\frac{k_{t}}{g_{f}|k_{r}|}dr.
\end{equation}
Therefore, the specific intensity $I(\upsilon_{obs})$ with infalling spherical accretion can be re-written as
\begin{equation}
\label{3-2-6}
I_{f}(\upsilon_{obs})\propto \int_{ray}\frac{g_{f}^{3}k_{t}dr}{r^{2}|k_{r}|}.
\end{equation}

\par
Similar to static spherical accretion, the specific intensity with infalling spherical accretion $I_{f}(\upsilon_{obs})$ also reaching a peak rapidly with the increased impact parameter and finally decreases gradually to a minimum value. For different RF parameters/BH charges, the numerical results of $I_{f}(\upsilon_{obs})$ are plotted in figure 5. Comparing figure 3 shows that the specific intensity shape of infalling spherical accretion is similar to static spherical accretion. Nevertheless, the infalling spherical accretion has a sharp rise before reaching the peak, which means that the photons are quickly captured by BH and resulting in a halo when the impact parameter reaches $b_{ph}$.
\begin{center}
\includegraphics[width=6cm,height=5cm]{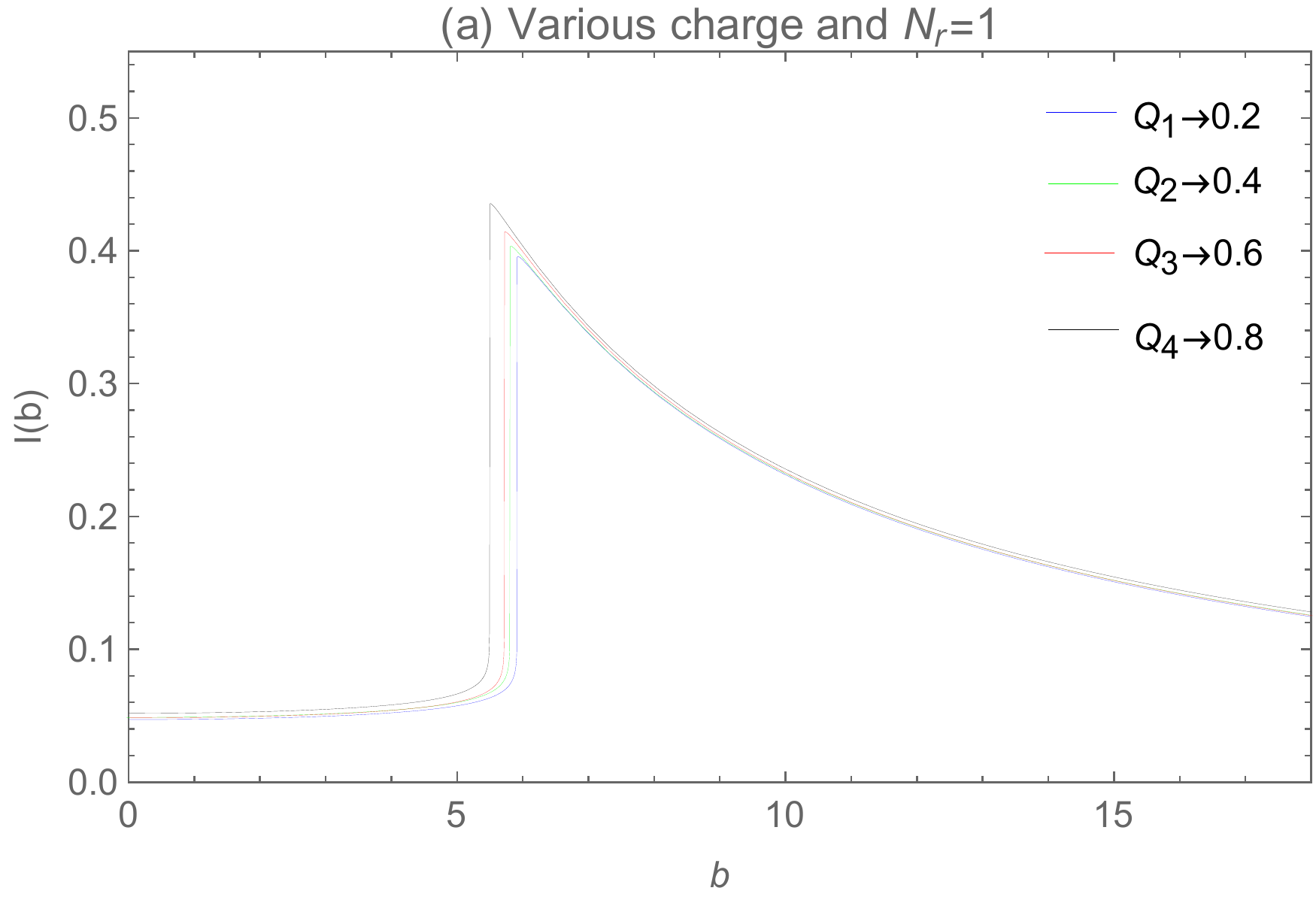}
\includegraphics[width=6cm,height=5cm]{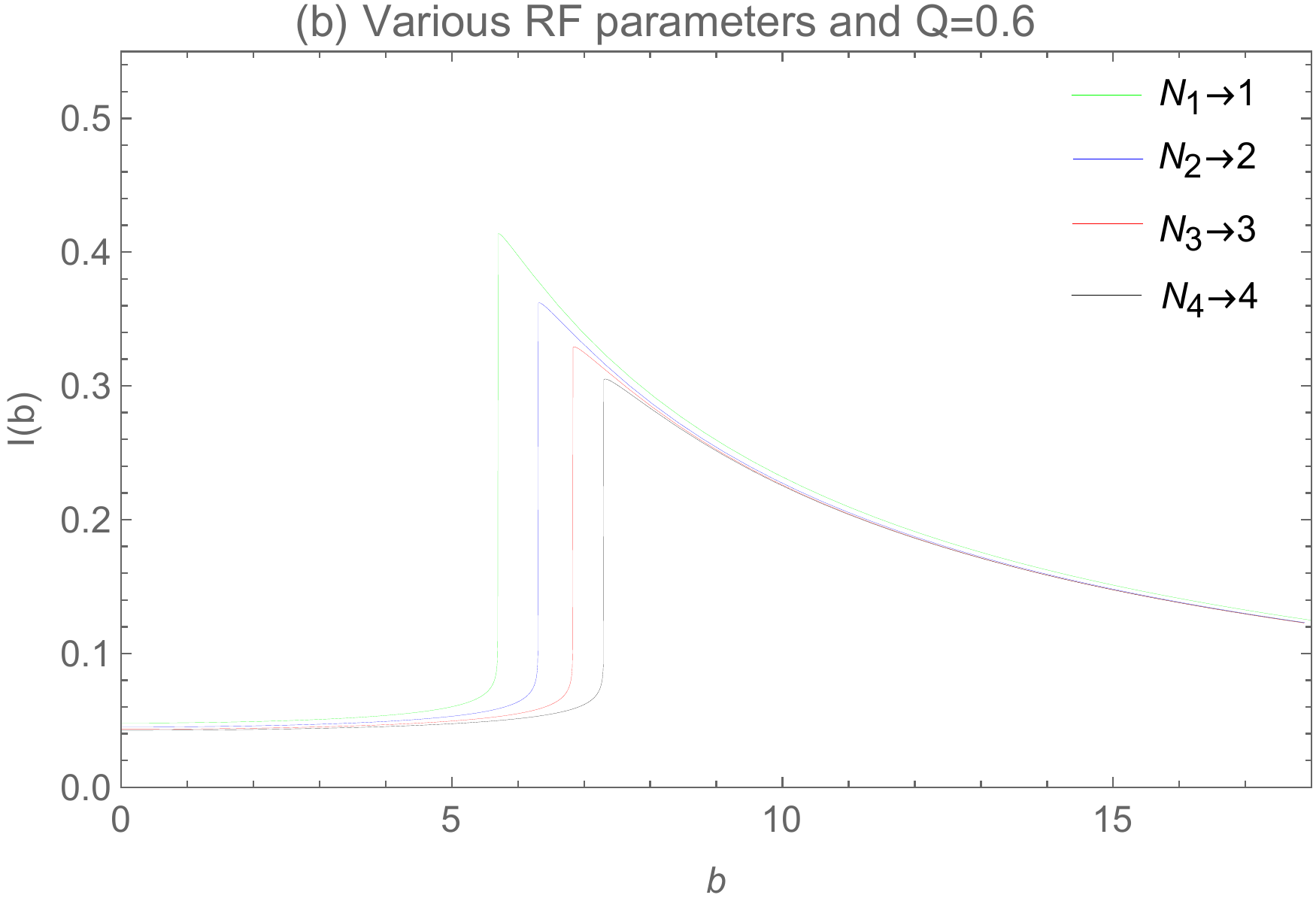}
\parbox[c]{15.0cm}{\footnotesize{\bf Fig~5.}  
Profiles of intensity $I(b)$ with infalling spherical accretion. {\em Panel (a)} -- Various charges $Q=0.2,0.4,0.6,0.8$ with $N_{r}=1, M=1$ and {\em Panel (b)} -- Various RF parameters $N_{r}=1,2,3,4$ with $Q=0.6, M=1$.}
\label{fig5}
\end{center}

\par
We also plot the BH shadow cast with infalling spherical accretion in figure 6. It is shown that the BH shadow size in infalling spherical accretion background is mainly affected by the RF parameter. The larger the RF parameter is, the larger radius of BH shadow will be. For the luminosity of the photon sphere, the more significant BH charge is, the photon sphere luminosity will be darker when the RF parameter is constant. While the BH charge is a constant, the luminosity of the photon sphere faintly increases gradually with the increase of the RF parameter.
\begin{center}
\includegraphics[width=6cm,height=6cm]{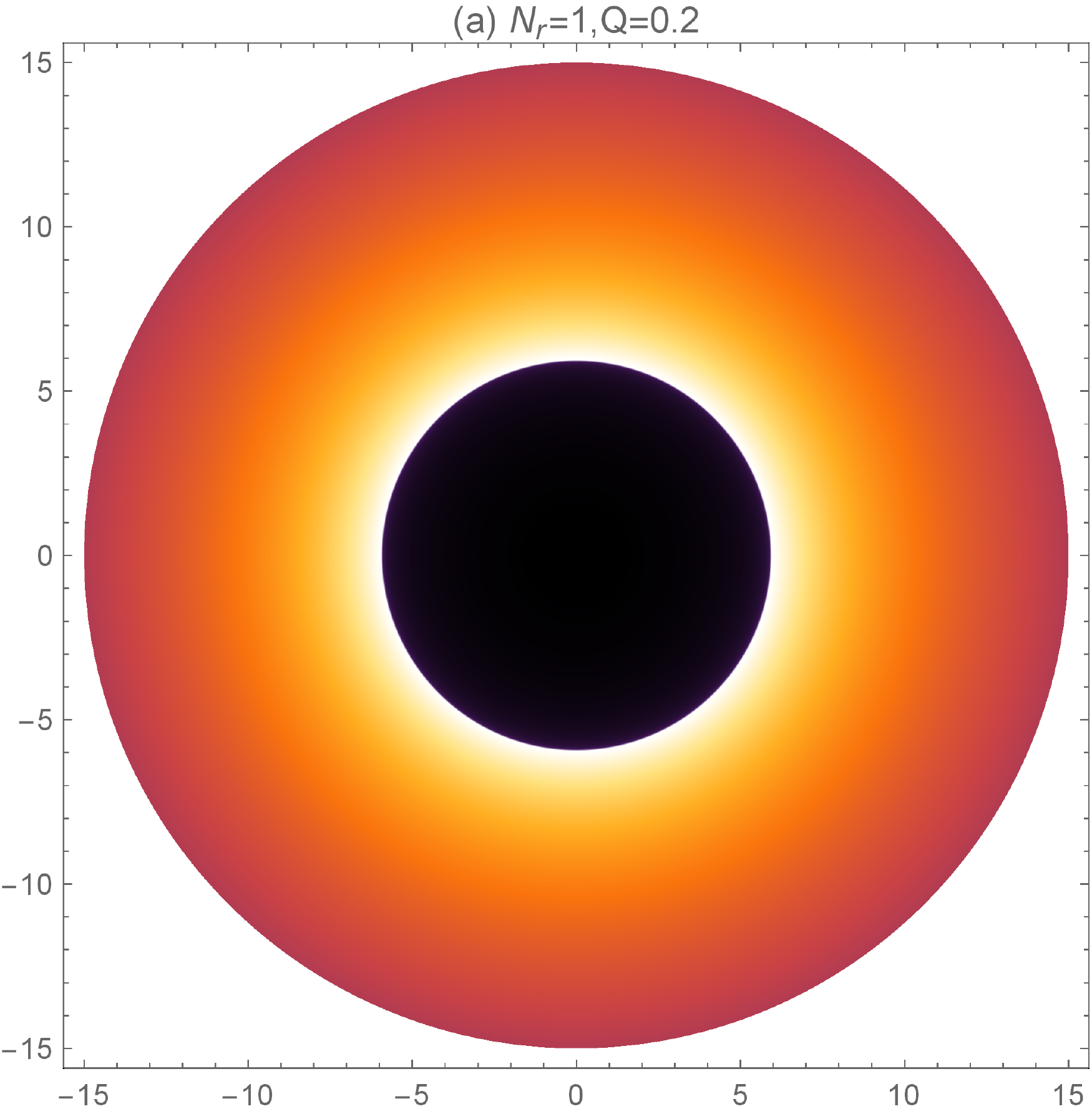}
\includegraphics[width=0.4cm,height=6cm]{fig0.pdf}
\includegraphics[width=6cm,height=6cm]{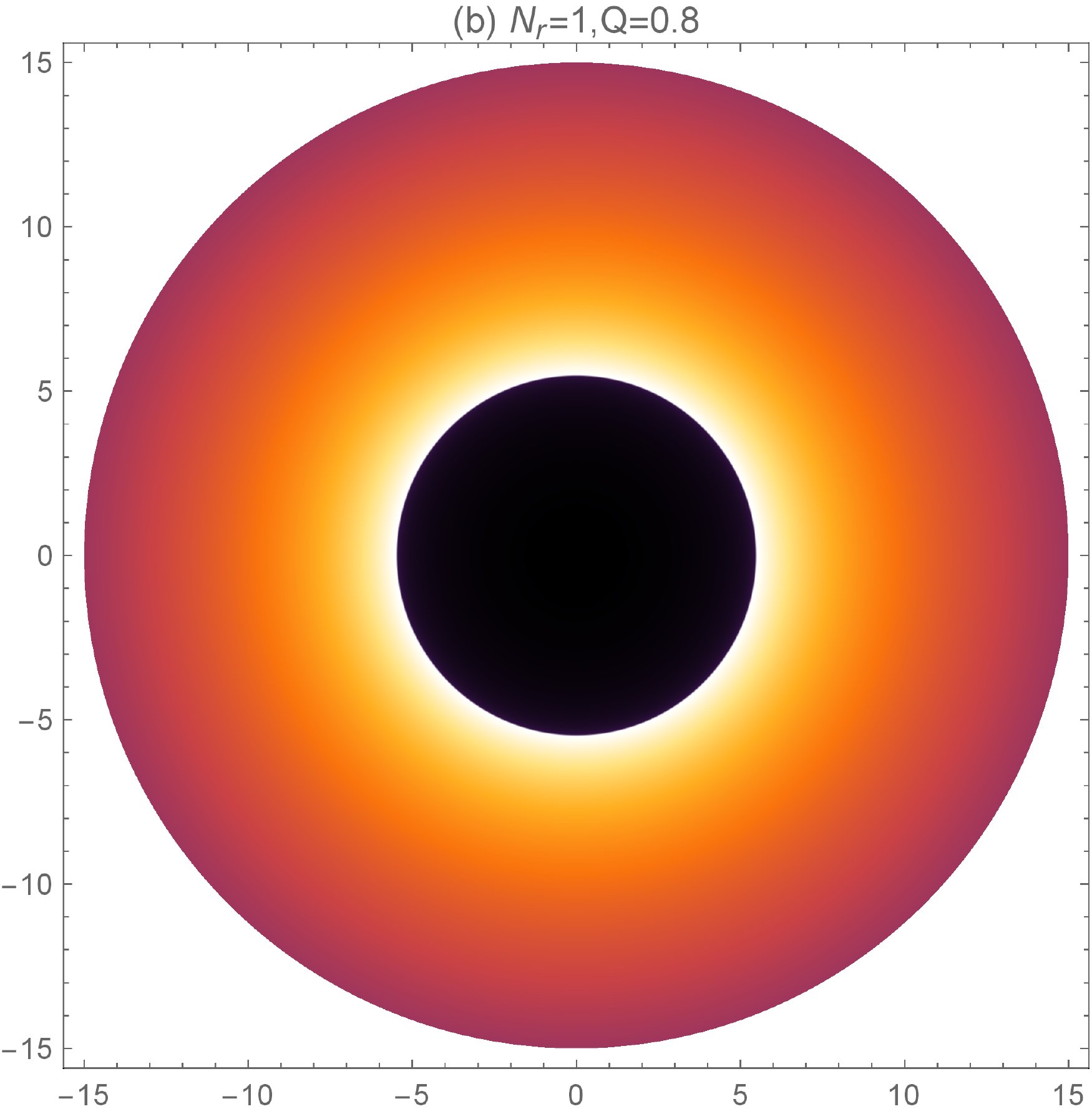}
\includegraphics[width=0.4cm,height=6cm]{fig0.pdf}
\includegraphics[width=6cm,height=6cm]{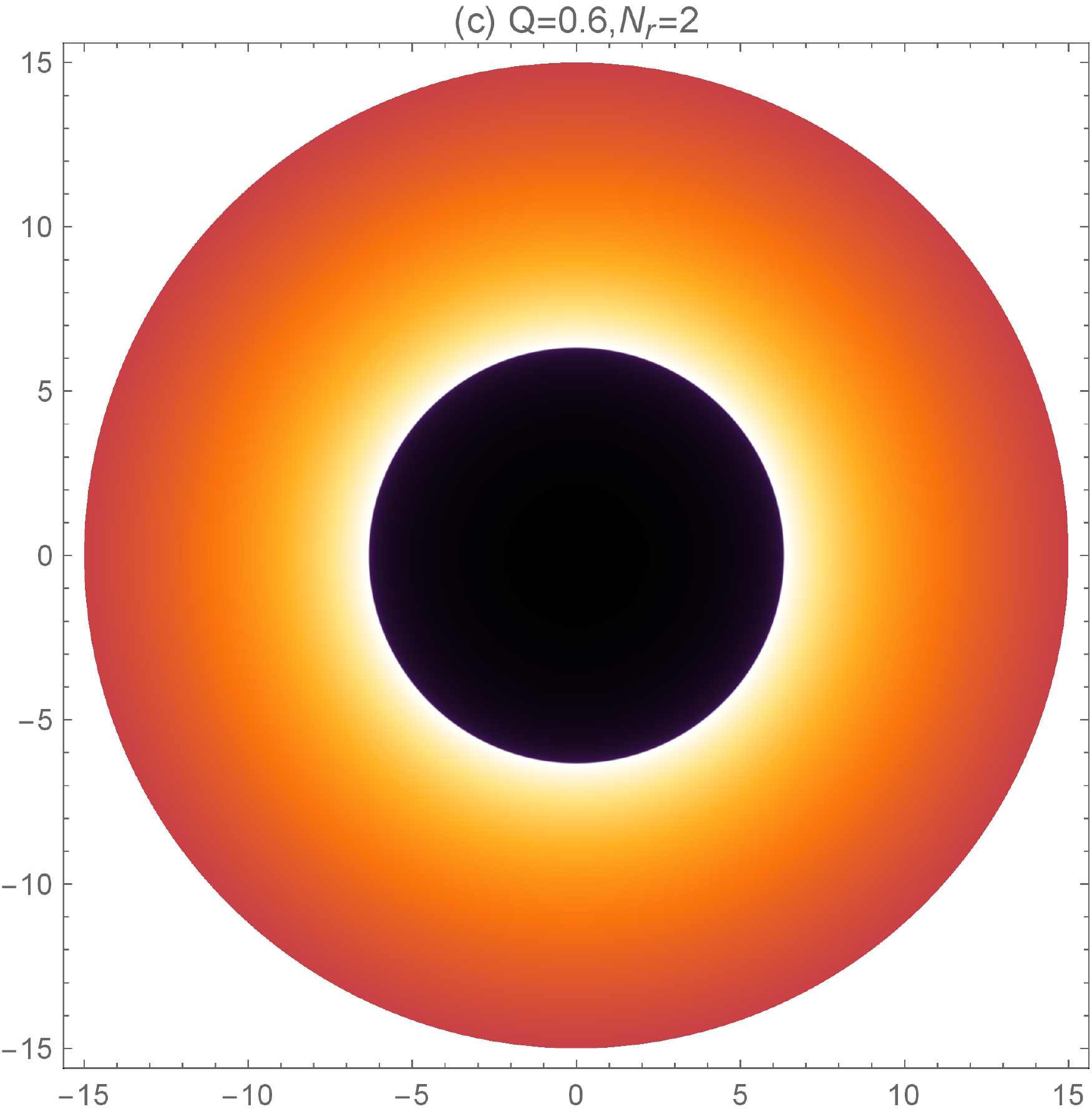}
\includegraphics[width=0.4cm,height=6cm]{fig0.pdf}
\includegraphics[width=6cm,height=6cm]{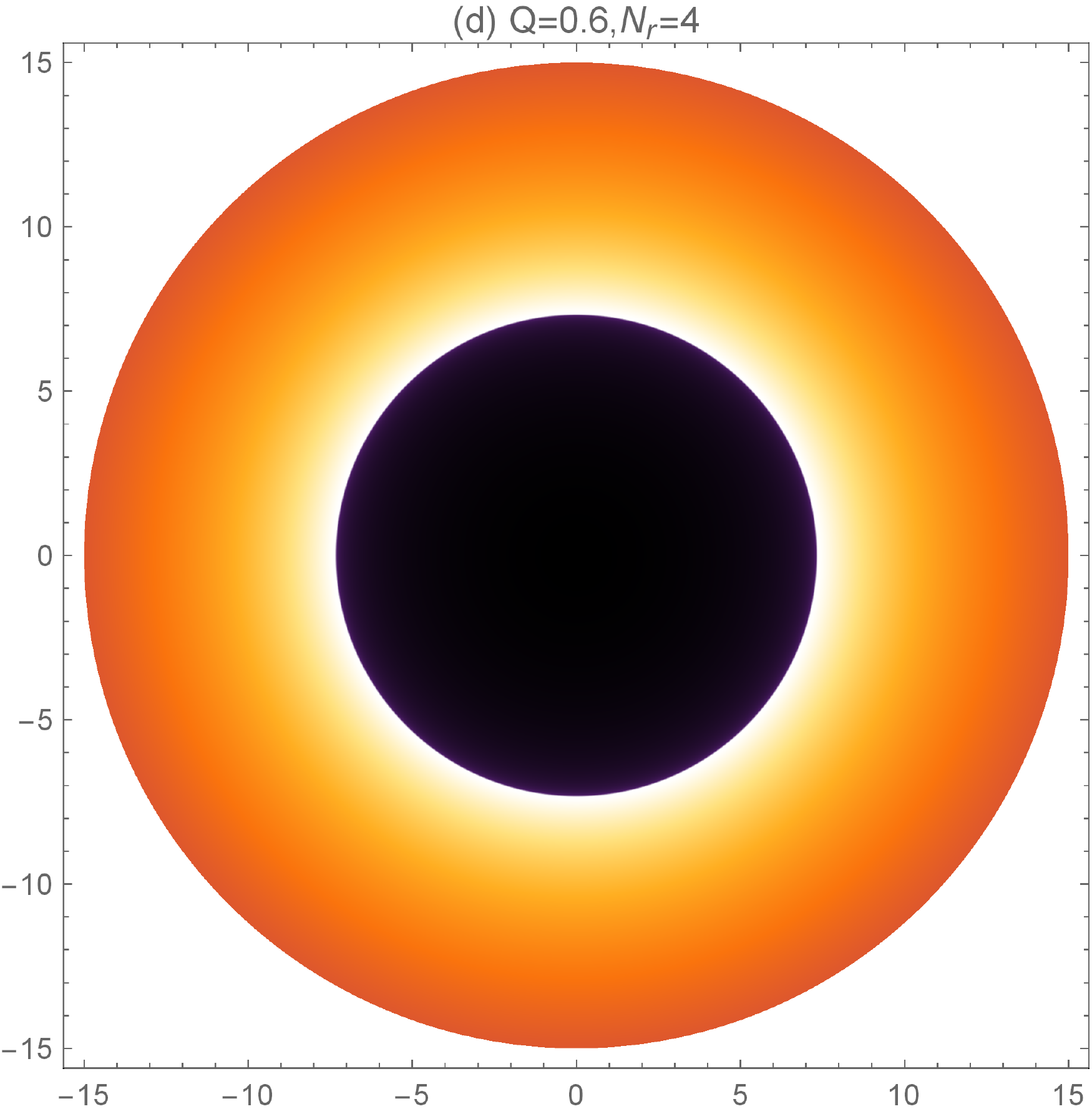}
\includegraphics[width=0.4cm,height=6cm]{fig0.pdf}
\parbox[c]{15.0cm}{\footnotesize{\bf Fig~6.}  
The BH shadows and photon spheres cast with infalling spherical accretion in (x,y) plane. {\em Panel (a, b)} -- Various charges $Q=0.2,0.8$ with $N_{r}=1, M=1$ and {\em Panel (c, d)} -- Various RF parameters $N_{r}=2,4$ with $Q=0.6, M=1$.}
\label{fig6}
\end{center}

\section{Conclusion}
\label{sec:4}
\par
In this paper, the shadow and photon sphere of a charged BH surrounded by PFRF in Rastall gravity on different spherical accretions background is investigated, and further creatively analyse the effect of RF parameter/BH charge on BH shadow and photon sphere. We mainly discussed BH shadow size and photon sphere luminosity with static and infalling spherical accretions, respectively. When the BH charge is constant, the increase of RF parameter leads to the simultaneous increase of BH event horizon radius, shadow radius and critical impact parameter. The BH charge positively affects the effective potential, and the RF parameter harms the effective potential. As a result, the observers see the BH shadow of different sizes in various BH charges/RF parameters.

\par
Then, we investigated the luminosity of BH shadow and photon sphere of this BH with static and infalling spherical accretions. We calculated the specific intensity observed by a distant observer and plotted the image of specific intensity as a function of impact parameter in various BH charges/RF parameters, which is shown that the peak value of intensity increase with BH charge increase when the RF parameter is a constant, but the $b$ corresponding to the peak value decreases with an increase of BH charge. However, the RF parameter oppositely affects the critical impact parameter, and these are consistent with the results in table 1 and table 2.

\par
This BH shadow and photon sphere cast in the $(x,y)$ plane is shown in figure 4 and figure 6, showing that the shadow inner region is not entirely black. The photon sphere with the strongest luminosity is outside of BH shadow. The photon sphere image maximum luminosity is attenuated with an increase of BH charge when the RF parameter is fixed. The luminosity of the photon sphere faintly increases gradually with increased RF parameter for given the BH charge. Compared with the change of luminosity, the RF parameter has more influence on BH shadow size.

\par
Compared with static and infalling spherical accretions, we obtain: \textit{i)} the size and position of BH shadow and photon sphere non-change in case of static and infalling spherical accretions, which imply that BH shadow is a signature of space-time geometry; \textit{ii)} BH shadow with infalling spherical accretion is darker than that of static spherical accretion in the central region, which means that most of the photons in static spherical accretion unit volume are captured by BH; \textit{iii)} the photon sphere with infalling spherical accretion is brighter than a static one, showing a small number of photons accumulate in photon sphere for static spherical accretion; \textit{v)} the increase of RF parameter has a positive effect on the photon sphere luminosity for different spherical accretions, while the effect of BH charge is opposite. The reason as the radiation field existence, the tiny fraction of photons escape from BH, and photon sphere luminosity becomes more prominent with the enhancement of RF parameter. On the other hand, the Hawking radiation suggests that BH with a more significant charge has a more substantial surface gravity $\kappa$, making photons easy to capture by BHs. From the point of view of spherical accretion, the impact parameter decreases with the increase in BH charge, which means that photons of light have more kinetic energy and make photons not easily captured by BHs.

\par
In our subsequent work, we will study the shadow and photon sphere of this black hole on the optically thin and geometrically thin/thick disk-shaped accretion background, and order to have a fuller understanding of the appearance of this black hole by investigating the photon rings and the lensing rings \cite{36}.

\section*{Acknowledgments}
The authors would like to thank the anonymous reviewers for their helpful comments and suggestions, which helped to improve the quality of this paper. This work is supported by the National Natural Science Foundation of China (Grant No.11903025).

\clearpage

\end{CJK}
\end{document}